\documentclass[seceq]{ptptex}

\usepackage{graphicx}
\usepackage{wrapft}

\def\qbarq{\langle \bar q q\rangle} \let \qqbar=\qbarq
\def\qgq{\langle \bar q \sigma\cdot Gq\rangle}
\def\fpi{f_{\pi}}

\title{
SUMMARY%
}

\author{
Makoto \textsc{Oka}%
}

\inst{
Department of Physics, Tokyo Institute of Technology, Meguro, Tokyo 152-8551
}

\abst{
This is a personal summary of the workshop.
I overview the topics of the workshop and itemize what we learned at the workshop.
}

\begin{document}

\maketitle

\section{Introduction}

Chiral symmetry and  its spontaneous breaking is one of the most fundamental aspects of QCD dynamics.
Recent active investigations have confirmed  that the ground state of QCD contains quark-antiquark pair condensation, which 
breaks chiral symmetry, and that the symmetry is reflected upon properties of hadrons.
Spontaneous symmetry breaking is a general phenomenon in various fields of physics,
such as superconductivity, and it is known that increase of temperature or chemical potential (density)
may cause phase transitions so that the system changes from the ordered phase to a disorder phase
by symmetry restoration.  In QCD, similarly, the restoration of chiral symmetry is expected at a high 
temperature and/or a high density.  Yet, until recently access to such symmetry restored phase was
not possible experimentally.

A new accelerator facility, RHIC at BNL, has opened a door to extensive studies of
high temperature hadronic matter, which is supposed to form quark-gluon plasma with unconfined 
quarks and gluons.  New data are being analyzed to confirm the QGP phase and to study its property
and dynamics of the transition.

Hadronic matter at high baryon number density is another place for the search of chiral restoration,
but no complete picture has been developed either experimentally nor theoretically.
Experimentally, nuclear density is not controllable to make the transition happen in the laboratory.  
Although the compact star, i.e., neutron stars and  possible quark stars, is a relevant laboratory of 
the high density nuclear matter, available observables are rather limited.
Theoretically, the lattice QCD calculation for finite density has disadvantage, that is, 
the action of QCD is complex for a finite chemical potential and is not suitable for Monte Carlo 
simulations.

Nevertheless,  for a recent few years, we have seen some important and exciting developments
for the finite density QCD.  If I may name a few,
\begin{enumerate}
\item  Several experiments which aim to produce mesons inside nuclei have been carried out
successfully and their data suggest precursory phenomena occurring in nuclear matter.  An 
exciting example is energy shifts of deeply bound pionic states, which indicate
modified pion decay constant in the nuclear medium.
\item  Possible existence of a new phase of hadronic matter, color superconducting phase, was
pointed out.  It is expected in a region of high baryon density and low temperature, 
and has nonzero $\langle q^TCq\rangle$ condensate, which breaks color gauge invariance, 
while the chiral symmetry is restored{\cite{csc}}.  
\item At a density of a few times of normal nuclear density, condensation of mesons, i.e., $\pi$, 
and /or $K$ as well as bulk strangeness brought by strange baryons, i.e., $\Lambda$, $\Sigma$, 
$\Xi$ etc., are considered to play significant roles.
\item  Development of the lattice QCD has just started to allow realistic simulation of  finite density QCD.
\item  Observations of the compact stars, neutron stars, quark stars, \ldots, are now providing us with
important information on the equation of motions of high density nuclear matter.
\end{enumerate}
 After all, the ultimate goal of hadron physics is to draw and complete the phase diagram of QCD
(Fig. 1) on the plane of the temperature ($T$) and the chemical potential ($\mu$).   
One sees in the phase diagram that the high $\mu$ QCD has indeed nontrivial physics content richer 
than high $T$. Chiral restoration at high density is the critical part to explore such physics.
\begin{wrapfigure}{2}{8.5cm}
\centerline{\includegraphics[width=8.5cm]{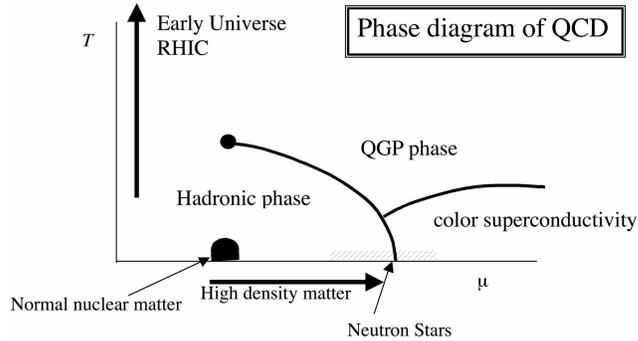}}
\caption{Expected phase diagram of QCD}
\label{fig:1}
\end{wrapfigure}

With all the above new developments in the high density QCD, it was quite timely to have this workshop, 
the YITP-RCNP Workshop on Chiral Restoration in Nuclear Medium at YITP.
In short, at this Workshop, we have seen a lot of materials 
on recent excitements in the area of finite density QCD.
We have seen the beginning of a new era, which is to develop the ``condensed matter physics'' in QCD.  
Obviously we have just started a step forward and this Workshop will be memorized as a milestone.

In the Workshop, we had about fifteen talks on theory and fifteen on experiments.
These numbers show that we have intimate communications and collaborations between experimentalists 
and theorists.  It is clear that such relation  is the key of recent developments in this field and will be as well
critically important in future.
It should also be pointed out that the research is international and naturally involves many countries 
in the world.  

\section{Overview}
Teiji Kunihiro, in his introductory remark, set the subjects of the workshop.
He itemized the following six topics (though not in this order)
\begin{itemize}
\item Suppression of the pion decay constant $f_{\pi}$ in medium.
\item Softening of the scalar spectral function.
\item Medium effects on the vector spectral function.
\item Possible deeply bound kaonic state in light nuclei.
\item Heavy quark hadrons in nuclear medium.
\item Baryons and chiral symmetry.
\end{itemize}
In all of these areas, new theoretical ideas as well as new experimental data were presented 
in the Workshop.  
I will briefly review some of (not all) the talks related to these topics in order.
Please refer to the individual report for the references, because I will omit them here.

\subsection{Suppression of the pion decay constant $f_{\pi}$ in medium}

Weise, in his opening talk, gave an overview on chiral dynamics at finite temperature 
and density.  According to the Gell-Mann-Oakes-Renner relation, the chiral condensate $\qqbar$ is
directly related to the pion decay constant $\fpi$.
Thus he argued that the thermodynamic behavior of $\qqbar$ can in principle be confirmed 
by measuring $\fpi$ in hot  and/or dense nuclear medium.

Recent observations of deeply bound pion states in heavy nuclei revealed significant suppression of
the binding energies of the 1s states in Pb and Sn isotopes.  
Weise argued that the missing repulsion can be explained by the medium modification of $\fpi$.
He showed that chiral symmetry, or the Weinberg-Tomozawa relation, predicts strong energy 
dependence of the zero-momentum $\pi N$ scattering amplitude in the isovector $t$-channel.  
This energy dependence is translated  as a wave function renormalization factor
in the energy-independent pion optical potential, 
$$ 2m_{\pi} U = \left[{\Pi (\omega, \vec q=0)\over
 1- {\partial\Pi\over \partial\omega^2}  } \right]_{\omega=m_{\pi}}+ \hbox{$p$-wave parts}
$$
and thus enhances the isovector component of the optical potential.  
This is equivalent to the reduction of $\fpi$ in nuclear medium{\cite{Weise}}.

Hirenzaki explained how such deeply bound mesonic states can be produced and showed how
they analyzed the energy shifts of deeply bound pionic states in heavy
nuclei, using the (energy-independent) optical potential model.
Suzuki presented their new analysis of the binding energies of the 1s atomic states of
deeply bound pionic atoms in heavy nuclei (Pb, Sn).  He showed that their analysis indicates
$$\fpi^*(\rho_0) \sim (0.63\pm 0.07) \times \fpi  .$$

During the discussion, Seki pointed out that the parameterization of the optical potential may have
ambiguities, especially in the coefficients of the momentum dependent term $b_0$, which 
is known to have a strong correlation with the two-body term $B_0$, and therefore has density
dependence.  
Then we should be cautious not to be too precise in making quantitative statements.
Nevertheless, the experimental data and the current analyses based on the in-medium
chiral dynamics indicate exciting possibility that we really ``measure" the chiral order
parameter as a function of density.

\subsection{Softening of the scalar spectral function}
Kunihiro stressed in his talk the roles of the scalar meson, $\sigma$, as the chiral
partner of the pion, the Nambu-Goldstone mode.  He showed that at finite density (and temperature)
the scalar correlation function is modified so that the $\sigma$ mode becomes softer, but at the same
time its strength accumulates and grows at above the threshold of two pions.
Ishida gave a review of the present status of the scalar meson spectroscopy, a convincing view
of the low mass $\sigma$ and the other members of SU(3) octet scalar mesons.
Cabrera also discussed possible medium modification in the photoproduction of two pions from nuclei.

As for experimental data, Grion showed that recent CHAOS data from the $(\pi, \pi\pi)$ reaction on nuclei 
show significant $\pi^+\pi^-$ enhancement at low energy.
Similar mass dependences were reported in the photoproduction of $\sigma\to\pi\pi$ experiments
at Mainz by Messchendorp, and SPring8 by Shimizu, while the Crystal Ball group, represented by Staudenmaier,
did not see $\sigma$ enhancement either for  the nucleon target nor the nuclear target.

Kunihiro also showed that the softening of the scalar mode
can be accounted for also in the nonlinear sigma model,
where the medium effect comes in the effective lagrangian from the term,
$$ -{g\fpi\over 2m_{\sigma}^2}\bar N N {\rm Tr} \left( \partial_{\mu} U\partial^{\mu} U \right)$$
and therefore corresponds exactly to the wave function renormalization of the pion{\cite{Jido}}.

Ericson pointed out that the scalar susceptibility, $\chi_s$, shows
a strong enhancement in nuclear medium.  
$$ \chi_s(\rho_0) \sim (5-6) \times \chi_s(0)\,.$$
This enhancement is caused by the softened $\qbarq$ fluctuation.
On the other hand, the pseudoscalar susceptibility, $\chi_{ps}$, is reduced
in medium, and thus under the chiral symmetry restoration they coincide with each other as is
expected from the symmetry.  
A similar degeneracy will happen in the vector and axialvector correlations.

\subsection{Medium effects on the vector spectral function and Heavy quark hadrons}
The temperature and density dependences of the vector spectral functions have been under 
intense investigation both theoretically and experimentally.
Hatsuda showed the vector (as well as the nucleon) spectral function calculated 
by the maximal entropy method (MEM) from the lattice QCD data.  
They studied temperature dependence of the spectral function and 
showed that the vector peak exists even above the critical temperature $T_c$.

Stachel, and En'yo presented electron-positron spectra at the CERES and KEK experiments, respectively.
Both show anomalous enhancement, 
which suggests the medium modification of the $\rho$ meson spectral
function.
An important question raised by Stachel is how big is the $\rho^2$ term in  $\qbarq_{\rho}$,
$$ -\qbarq_{\rho} =  -\qbarq_0 (1- C\rho/\rho_0+D (\rho/\rho_0)^2 ) \, .$$
The CERES data suggest that a significant $\rho^2$ term exists, 
but we do not have complete picture yet.

Kienle summarized the various experimental projects at GSI, probing the behaviors of various
mesons in hot and dense medium, i.e., deeply bound $\pi^-$, $\eta$, $K^-$, $\omega$ states,
the dilepton spectra, $D$ meson production and so on.
The spectrum of heavy quark mesons and baryons, such as $J/\psi$ or $D$ mesons, was also the 
subject of the talks by Lee, Blaschke and Tsushima.

\subsection{Possible deeply bound kaonic state and Strangeness in medium}
The strangeness is unique in QCD because $m_s$ is comparable to $\Lambda_{\rm QCD}$.
Thus, its dynamics is sensitive to the nonperturbative QCD and is not trivially determined by chiral symmetry.
For instance, the behavior beyond chiral perturbation theory may allow us to interpolate
the chiral effective theory and the perturbative QCD regime.

Behavior of the kaon in nuclear medium attracts a lot of attention recently.  
In this workshop, Akaishi showed their recent calculation of deeply bound kaon states in light nuclei.
According to their picture, the isoscalar kaon-nucleon interaction is strongly attractive so that the
$K^-$ imbedded in nuclei attracts surrounding nucleons making a dense $\bar K$-nucleus bound
state.  Although validity of such calculation is under a heated discussion,{\cite{Gal}}
experimental efforts for confirmation of this exciting suggestion are encouraged. 
Some ongoing experiments were reported by Itahashi, Cargnelli and Kishimoto.  
Kishimoto claimed that their preliminary data of the $^{16}$O$(K^-, N)$ reaction  at AGS are indeed consistent with a deep attractive potential for $K^-$.
Hotta discussed photoproduction of $\Lambda(1405)$.

\subsection{Lattice QCD at finite density and Vector manifestation of chiral symmetry}
It is notorious that the lattice QCD is not properly applied to finite baryon density, or chemical potential,
because the measure of Monte Carlo integral becomes complex.  Recently, some techniques to avoid
this difficulty have been invented{\cite{LQCDrho}}.  
Nakamura, in this workshop, explained their efforts to simulate
SU(2) lattice at finite chemical potential, which circumvents the above mentioned difficulty.
He showed that the $\rho$ meson mass decreases and the masses of $\rho$ and $\pi$ are interchanged 
as the chemical potential increases.

Rho argued that this behavior of the vector meson mass may be attributed to a non-standard scenario of
chiral symmetry restoration, vector manifestation.  Harada explained how the vector manifestation is
concluded from their analysis based on the renormalization group method in QCD at finite temperature 
and/or density.  In this new manifestation, the longitudinal component of the vector meson becomes the chiral
partner of the NG boson, and in the chiral limit, it is decoupled from the transverse part
as the mass of the vector meson goes to zero.  
Then the axialvector and pseudoscalar correlations behave at  $\mu \to \mu_c$ as
\begin{eqnarray*}
 \bar q \gamma^5\gamma^{\mu} q &\sim& F_{\pi} \pi  q^{\mu}\cr
\bar q \gamma^5 q &\sim& G_A A_{\mu} q^{\mu}
 \end{eqnarray*}
 with nonzero $F_{\pi}$ and $G_A$.
This provocative suggestion caused a heated discussion.
Some of the objections are based on the points that their RG analysis does not take the axialvector meson 
into account, and that this restoration may belong to a wrong universality with extra massless degrees of
freedom.

\subsection{Baryons and chiral symmetry}
Chiral symmetry gives rather strong constraints on the meson spectrum and properties, while the baryon
is usually regarded as heavy matter field not much to do with chiral symmetry.  Hosaka argued, however, 
that the linear chiral representation for the nucleon and its excited states allows an interesting 
non-standard way of assignment, that is, the mirror assignment of chiral symmetry to negative
parity nucleon excited states.  
Chiral transformation, $L\in SU(N_f)_L$, $R\in SU(N_f)_R$, on the baryons, $B_1$ and $B_2$,
are given by
\begin{eqnarray*}
 B_{1L} &\to& L B_{1L} \qquad  B_{1R} \to R B_{1R} \cr
 B_{2L} &\to& R B_{2L} \qquad  B_{2R} \to L B_{2R} 
 \end{eqnarray*}
 for the mirror baryons.
If the mirror chiral partner of the nucleon is an excited nucleon 
resonance, they should show characteristic features in medium due to the partial restoration of
chiral symmetry.
It was also pointed out that the mirror baryons have opposite signs of the axial charge, which 
may be confirmed experimentally.
This scheme allows us to have a chiral symmetric mass term in the effective lagrangian,{\cite{Jido_Hosaka}}
$$  L_{\rm mass} \sim -m_0 (\bar B_2 B_1 + \bar B_1 B_2) \, .$$
Such a mass $m_0$ is independent of chiral condensate and thus survives in the chiral restoration,
$\qbarq \to 0$.

Kasagi discussed photoproduction of $\eta$ on the nucleon and nuclei,
and the in-medium behavior of the $N^*(1535)$ resonance state, the lowest negative parity 
excitation of the nucleon,  which may confirm the mirror baryon assignment.

\section{Conclusion}

Most of the participants of the workshop seemed to agree in the following points.

\smallskip\noindent (1) Chiral symmetry will be restored at $T\to T_c$ and $\mu\to\mu_c$.

\smallskip\noindent (2) The order parameter is the quark condensate, which behaves at finite density as
$$\qbarq \sim\fpi \sim (1-C\rho/\rho_0)$$ 
with $C\approx 0.2 -0.4$, although $\rho^2$ term may also be significant.

I also pointed out that the other order parameter of chiral symmetry breaking is also
important to study.  The quark-gluon mixed condensate is one of the most important, as
it is the condensate with the next-lowest dimension, $D=5$.
According to the QCD sum rule analyses, $\qgq$ is as large as the (1GeV)$^2\times\qbarq$.
Interesting question to ask is whether the behaviors of $\qgq$ condensate at finite density and/or temperature 
are the same or similar to the $\qbarq$.  At finite $T$, it is a question to be answered in the lattice QCD, 
which is underway.{\cite{Doi}}

\smallskip\noindent (3) A light scalar meson $\sigma$ a.k.a. $(\pi\pi)_ {I=0, J=0}$, the
chiral partner of the Nambu-Goldstone boson $\pi$, will signal
chiral restoration by spectral softening in the course of $\rho = 0\to \rho_0 \to \rho_c$. 

\smallskip\noindent (4) Meson bound states in nuclei are good probes of the finite density chiral 
dynamics.  Suppression of $\fpi$ extracted from the deeply bound pion state is 
a benchmark example.

\smallskip\noindent (5) Various other probes should be pursued, such as, $\eta$ bound states, 
$J/\psi$ and $D$ production, nucleon  resonance excitations in nuclei.

\smallskip\noindent (6) Strange hadrons in nuclei. especially, kaon bound states, hyperon properties in hypernuclei, 
and  $\Lambda$ excited states produced in nuclei are good probes of chiral dynamics as well.

\bigskip
In conclusion, after much intensive discussion, we still have many unsolved problems.
All the participants have got many assignments, and I hope many of the questions
are solved before the next workshop.

\section*{Acknowledgements}
I would like to thank Dr. Teiji Kunihiro and Atsushi Hosaka for giving me opportunity to attend this workshop
and to give a summary.

%


\begin{thebibliography}{99}
\bibitem{csc} M. Alford, K. Rajagopal, F. Wilczek, \PLB{422 ,1998,247} \\
R. Rapp, T. Schaefer, E.V. Shuryak and M. Velkovsky, \PRL{81,1998,53}.
\bibitem{Weise} W. Weise \NPA{690,2001,98}.
\bibitem{Jido} D. Jido, T. Hatsuda, T. Kunihiro, \PRD{63,2001,011901}.
\bibitem{Gal} A. Gal, \NPA{691,2001,268c}
\bibitem{LQCDrho} Z. Fodor, S.D. Katz, \PLB{534,2002,87}.
\bibitem{Jido_Hosaka}  C. DeTar and T. Kunihiro, \PRD{39,1989,2805} \\
D.Jido, M. Oka, and A. Hosaka, \PTP{106,2001,873}, \PTP{106,2001,823}.
\bibitem{Doi} T. Doi, N. Ishii, M. Oka, and H. Suganuma, hep-lat/0211039.
\end{thebibliography}
\end{document}